\documentstyle[12pt, epsf]{article}
\newfont{\bbb}{msbm10}

\newcommand{\til}{\mbox{$\tilde{x}_{2}$}}
\newcommand{\barra}{\mbox{$\bar{x}_{2}$}}
\newcommand{\reals}{\mbox{{\bbb R}}}

\newtheorem{Prop}{Proposition}
\newtheorem{Theo}[Prop]{Theorem}
\newtheorem{Lemma}[Prop]{Lemma}
\newtheorem{Def}[Prop]{Definition}
\newtheorem{Cor}[Prop]{Corollary}

\newtheorem{Example}[Prop]{Example}

\begin{document}

\bibliographystyle{plain}

\title{Hybrid Dynamics of Two Coupled Oscillators that can Impact a Fixed Stop}

\author{Andr\'{e} Xavier C. N. Valente\\ 
Division of Engineering and Applied Sciences \\
Harvard University, Cambridge, MA 02138, U.S.A.\\
E-mail: andre@deas.harvard.edu\\
\and N. H. McClamroch\\
Department of Aerospace Engineering\\ 
University of Michigan, Ann Arbor, MI 48109, U.S.A\\
E-mail: nhm@umich.edu\\
\and Igor Mezi\'{c} \\
Dept. of Mechanical and Environmental Engineering\\
and Dept. of Mathematics\\
University of California, Santa Barbara, CA 93106, U.S.A.\\
E-mail: mezic@engineering.ucsb.edu}

\maketitle

\begin{abstract}
We consider two linearly coupled masses, where one mass can have inelastic impacts with a fixed, rigid stop.  This leads to the study of a two degree of freedom, piecewise linear, frictionless, unforced, constrained mechanical system. The system is governed by three types of dynamics: coupled harmonic oscillation, simple harmonic motion and discrete rebounds. Energy is dissipated discontinuously in discrete amounts, through impacts with the stop. We prove the existence of a nonzero measure set of orbits that lead to infinite impacts with the stop in a finite time.  We show how to modify the mathematical model so that forward existence and uniqueness of solutions for all time is guaranteed. Existence of hybrid periodic orbits is shown. A geometrical interpretation of the dynamics based on action coordinates is used to visualize numerical simulation results for the asymptotic dynamics.
\end{abstract}

\section{Introduction}
Mechanical impact oscillators are ideal case studies for the investigation 
of the dynamical behaviors possible in systems
 governed by a mixture of continuous and discrete dynamics. 
 Guckenheimer and Holmes \cite{Guck} (pp. 104-116)
 discuss the dynamics of a ball bouncing on a sinusoidally forced vibrating table.
  In \cite{Vasu} Salapaka et al. study a variation of this system, where
an unforced harmonic oscillator interacts with a sinusoidally vibrating table.
 A one degree of freedom periodically forced oscillator with a stop restricting its motion is studied
 by Shaw and Holmes \cite{Shaw}.
 Hindmarsh and Jefferies \cite{Hindmarsh}
 analyze the impact oscillator for the case in which the position of the
 wall is offset from the rest point of the oscillator. 
 New types of bifurcations that arise in connection with grazing impacts have been extensively studied;
 see Whiston \cite{Whiston}, Nordmark \cite{Nordmark}, Chin et al. \cite{Chin} and Budd and Dux \cite{Budd}.
 For a study of a two degrees of freedom impact oscillator, see the work by J. Shaw and S. Shaw \cite{Jing}. 
  In \cite{Engleder}, Engleder et al. investigate, numerically and experimentally, a non-smooth system consisting of two adjacent beams that can impact each other, vibrate independently or vibrate as a single unit.
 The paper by Stewart \cite{Stewart} provides a good survey of fundamental problems arising in the modelling of hybrid systems. We refer the reader interested in a more in depth mathematical discussion on the modelling of discontinuous differential equations to Deimling \cite{Deim}, Filippov \cite{Filippov} or Monteiro \cite{Monteiro}.

\begin{figure}[h] 
\vspace{.2in}
\epsfysize=1.6 in
\centerline{\epsfbox{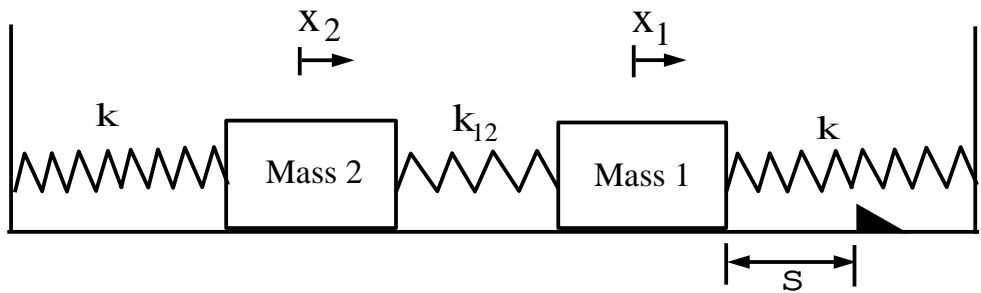}}
\vspace{-.1in}
\caption{The two degree of freedom impact oscillator.} \label{main figure}
\vspace{.2in}
\end{figure}

In this paper we analyze the dynamics of the two degrees of freedom coupled oscillator constrained by a stop shown in Figure \ref{main figure}. There is no external forcing. The springs are linear and frictionless. Mass one and mass two interact only via the spring $k_{12}$, as in a classical coupled oscillator. The sole impacts in the system are therefore those of mass one with the stop. Collisions with the stop are treated using a coefficient of restitution model, i.e., the velocity of the mass just after an impact is assumed to be always a constant fraction $\varepsilon$ of its velocity just before the impact. Collisions are assumed to be inelastic, with $0\!<\!\varepsilon\!<\!1$. The stop is offset a positive distance away from the equilibrium position of mass one. 
 
This is an instance of a hybrid dynamical system, also known as a non-smooth dynamical system. There exist three distinct types of motion. If mass one is not touching the stop then we have the usual two coupled harmonic oscillators. If mass one hits the stop, then there is an instantaneous rebound, governed by the 
coefficient of restitution model.
Thirdly, mass one could be constrained to remain against the stop by the force acting 
on it, leaving mass two moving as a single harmonic 
oscillator.

An interesting feature of the dynamics is the presence of orbits that lead to 
infinite impacts of mass one with the stop in a finite time.
 We prove the existence of such orbits and we show that they can be continued from the limit point they were approaching, in order to guarantee existence of solutions for all time.
 Another interesting feature of this system is the way energy is dissipated.
Because the springs are frictionless and the coefficient of restitution is less than unity, energy is dissipated only at impacts of mass one with the stop.
Furthermore, at such an impact, energy is instantaneously reduced by a discrete amount, since the speed of the impacting mass is reduced by a discrete amount.

The paper is divided into two parts, {\em Modelling} and
 {\em Dynamics of the System}. The {\em Modelling} part is concerned with rigorously establishing forward existence and uniqueness of solutions for all time. We start by giving an exact description of the system and its governing dynamics. In section \ref{infrebsec} we prove the existence of a nonzero measure set of orbits that lead to infinite rebounds in a finite time. Then, in section \ref{analyticalcont}, we show how to modify our mathematical model to ensure existence and uniqueness of solutions for all time. These two sections are the more mathematical ones of the paper. Readers whose interest lies more in the general dynamical behavior of our system may consider skipping them. Conversely, readers whose interest is the problem posed by infinite rebounds and how to possibly deal with that problem, may consider skipping the last part, {\em Dynamics of the System}, as this part concentrates on the dynamics of our particular system. 
 In the {\em Dynamics of the System} part, we present some of the more interesting dynamical behaviors of the system.
Section \ref{PHO section} is devoted to hybrid periodic orbits, a surprising form of periodic behavior where the system alternates between coupled harmonic motion and simple harmonic motion with mass one pressed against the stop. Section \ref{Action section} gives a geometrical interpretation of the system dynamics based on action coordinates. Using it, we also visualize some numerical simulation results.

\section{Modelling}

\subsection{The basic model}

Figure \ref{main figure} shows the system under study and the coordinate system used. 
 The origin of the $x_{1}$ and $x_{2}$ coordinates in Figure \ref{main figure} correspond, respectively, to the 
equilibrium positions of masses one and two if the stop were not present. 
 Constants $k$ and $k_{12}$ denote respectively
the outer and middle spring constants as shown in the figure. 
 The forces on mass one and mass two are denoted by $F_{1}$ and $F_{2}$.
 For simplicity, we let both masses be unity throughout the paper.
The position of the stop, denoted by s, 
is measured with respect to the reference defined by $x_{1}=0$.
 At the equilibrium position mass one is not leaning against the stop, that is, s is 
assumed to be strictly positive. 
 Also, in all cases it is assumed that mass one is located to the left of the stop.
 
Depending on the value of the state variables, $(x_{1},v_{1},x_{2},v_{2})$,
 where $v_{1}=\dot{x}_{1}$ and $v_{2}=\dot{x}_{2}$, a different set of equations governs 
the evolution of 
the motion. Namely, at any time instant one of the following three types of dynamics holds:

1. {\em Coupled Oscillator Dynamics (COD):} The coupled harmonic oscillator dynamics governs
the system when mass one is not against the stop.

2. {\em Rebound Map (R):} When mass one hits the stop with a nonzero velocity, an instantaneous
rebound occurs. The rebound is modeled using a coefficient of restitution model.
The coefficient of restitution is assumed to be strictly between zero and one.  

3. {\em Contact Dynamics (CD):} If at some instant mass one reaches the stop with zero velocity,
 then the force on it may constrain it to remain stuck against the stop. 
During such a period, the second mass undergoes simple harmonic motion. 

Formally, we divide the phase space into three regions and assign to each region one of the 
aforementioned dynamics. This is laid out in chart \ref{bigchart}.

To simplify the way we denote regions in the phase space, we will use
an informal notation for sets. For example $\{s\!=\!10,x_{2}\leq c\}$
 stands for
$\left\{ {\bf x}=(x_{1},v_{1},x_{2},v_{2}) \right.$
 $\left. \in S: x_{1}=10\wedge x_{2}\leq c \right\}$
, where S is the phase space. 
\\
\\
\\
{\bf Chart \ref{bigchart}.} \label{bigchart}
\vspace{.22in}

Phase Space: $S\equiv \{ (x_{1},v_{1},x_{2},v_{2}) \in \reals^{4}\ |\ x_{1}\leq s\}$
\vspace{.12cm}
\begin{enumerate}
\item
Coupled Oscillator Dynamics {\em (COD)}

Region: $\{x_{1}\!<\!s, v_{1},x_{2},v_{2}\} \cup
\{s,v_{1}\!<\!0,x_{2},v_{2}\} \cup
\{s,v_{1}\!>\!0,x_{2},v_{2}\} \cup
\{s,0,x_{2}\leq \tilde{x}_{2},v_{2}\}$, 

where $\tilde{x}_{2}= \left( \frac{k+k_{12}}{k_{12}} \right) s$

Entrance Surface: $\{s,v_{1}<0,x_{2},v_{2}\} \cup \{s,0,\tilde{x}_{2},v_{2}< 0\}$

Exit Surface: $\{s,v_{1}>0,x_{2},v_{2}\} \cup \{s,0,\tilde{x}_{2},v_{2}>0\}$

Dynamics:
$
\label{CODeqns}
\left\{ \begin{array}{l}
\dot{x}_{1}=v_{1} \\
\dot{v}_{1}= -(k+k_{12})x_{1}+k_{12}x_{2}\\
\dot{x}_{2}=v_{2} \\
\dot{v}_{2}=-(k+k_{12})x_{2}+k_{12}x_{1}\\
\end{array} \right.  
$

\vspace{.12cm}
\item
Contact Dynamics {\em (CD)}

Region: $\{s,0,x_{2}>\tilde{x}_{2},v_{2}\} \cup 
\{s,0,\tilde{x}_{2},v_{2}>0\} \cup
\{s,0,\tilde{x}_{2},v_{2}<0\}
$

Entrance Surface: $\{s,0,\tilde{x}_{2},v_{2}>0\}$

Exit Surface: $\{s,0,\tilde{x}_{2},v_{2}<0\}$

Dynamics:
$
\label{CDeqns}
\left\{ \begin{array}{l}
\dot{x}_{1}=0 \\
\dot{v}_{1}=0\\
\dot{x}_{2}=v_{2} \\
\dot{v}_{2}=-(k+k_{12})(x_{2}-x_{2eq}) \mbox{ where, } x_{2 eq}=\left( \frac{k_{12}}{k+k_{12}} \right) s

\end{array} \right.  
$

\vspace{.12cm}
\item
Rebound Map {\em (R)}

Region: $\{s,v_{1}>0,x_{2},v_{2}\} \cup \{s,v_{1}<0,x_{2},v_{2}\}$

Entrance Surface: $\{s,v_{1}>0,x_{2},v_{2}\}$

Exit Surface: $\{s,v_{1}<0,x_{2},v_{2}\}$

Dynamics:
$
\label{Reqns}
\left\{ \begin{array}{l}
x_{1}^{new}=x_{1}^{old} \\
v_{1}^{new}=-\varepsilon v_{1}^{old}\\
x_{2}^{new}=x_{2}^{old} \\
v_{2}^{new}=v_{2}^{old}\\
\end{array} \right.  
$
\end{enumerate}

\vspace{.4cm}

In the above chart, the Entrance and Exit surfaces of a region are, respectively, the sets through which orbits enter and leave that region. 
For the rebound map, 
one defines the entrance surface as the domain of the map and
 the exit surface as its range.

\begin{figure}[h]
\vspace{.2in}
\epsfysize=1.8 in
\centerline{\epsfbox{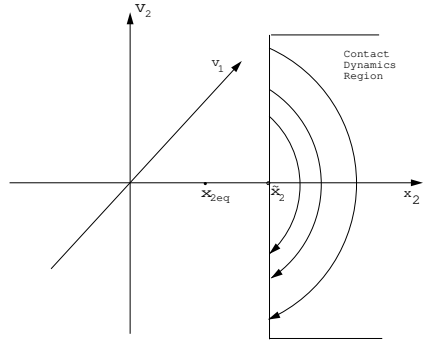}}
\caption{The $x_{1}\!=\!s$ hypersurface.}  \label{hypersurface}
\vspace{.2in}
\end{figure}

Let us see how the 
division of the phase space relates to the actual physical dynamics of our impact oscillator.
Clearly the region $\{x_{1}\!<\!s\}$ is
governed by COD. More interesting are the dynamics and division
of the phase space at the hypersurface $\{x_{1}=s\}$ which physically corresponds to mass one being in 
contact with the stop. This 3-D hypersurface is shown in Figure \ref{hypersurface}.

The $\{x_{1}\!=\!s, v_{1}\!>\!0\}$ region is the rebound entrance set. After a rebound, $v_{1}$ becomes negative and mass one leaves the stop, hence points in 
$\{x_{1}\!=\!s,v_{1}\!<\!0\}$ fall again under COD.
 Points on the plane $\{x_{1}\!=\!s, v_{1}\!=\!0\}$
 are reached when mass one decelerates towards the stop and reaches 
it with zero speed. 
In this case, if the force, $F_{1}$, on mass one is positive then mass one stays pressed against the stop, otherwise it does not.
 The critical position of mass two that yields $F_{1}\!=\!0$ when mass one is at the stop is $x_{2}\!=\!\tilde{x}_{2}$.
 The point $x_{2eq}$ is the equilibrium position of mass two when mass one is against the stop. Hence, it
is the center of the simple harmonic motion of mass two under CD.
 The continuous lines on Figure \ref{hypersurface} represent these CD orbit segments.
 Observe that the existence of the CD orbits moving on the $\{x_{1}\!=\!s\}$ surface precludes the use of this surface as a Poincar\'{e} cross section where one could define an implicit composite map to describe the global dynamics. 
\\

Now, one would like to formally establish forward existence and uniqueness of solutions.
It turns out that each of our three types of dynamics guaranteeing existence and uniqueness within
 their assigned regions is not enough to establish global existence and uniqueness of solutions 
for all time. In non-smooth systems, forward existence can be broken by orbits that lead to an infinite 
number of switches between different types of dynamics in
a  finite time, for these orbits are clearly undefined for times posterior to the occurence of
those infinite switches.
 Surprisingly, such orbits do occur in our system and in fact their set has nonzero measure.
 These are orbits that undergo infinite rebounds in a finite time, in other words, orbits
that switch an infinite number of times between {\em COD} dynamics
and {\em R} dynamics in a finite time.
 In the next section we will prove the existence of such orbits and that their set has nonzero measure.
Then, in section~\ref{analyticalcont}, we solve the existence problem through continuation of these orbits from the limit point they were approaching.
  We also rule out any other type of infinite switching in finite time, hence guaranteeing forward existence and uniqueness of orbits in the system.

\subsection{Proof of the existence of a nonzero measure set of orbits that lead to infinite rebounds in a finite time} \label{infrebsec}

This section is devoted to proving the theorem below.

\begin{Theo} \label{SuperProp}
There exists a nonzero measure set of orbits that lead to infinite rebounds in a finite time.
\end{Theo}

For reference, we group below all definitions and basic assumptions used in this section.

\begin{enumerate}
\item[] $F_{1}(t)\equiv$ net force on mass one.
\item[] $F_{2}(t)\equiv$ net force on mass two.
\item[] $\tilde{x}_{2}\equiv \left( \frac{k+k_{12}}{k_{12}} \right) s$, the position of mass two that results in 
$F_{1}\!=\!0$ when $x_{1}\!=\!s$.
\item[] $x_{2 eq}\equiv \left( \frac{k_{12}}{k+k_{12}} \right) s
$, the position of mass two that results in 
$F_{2}\!=\!0$ when $x_{1}\!=\!s$.
\item[] $v_{out_{n}} \equiv $ velocity of mass one just after the $n$th impact.
\item[] $v_{in_{n}} \equiv $ velocity of mass one just before the $n$th impact.
\item[] $\Delta t_{n} \equiv $ time interval between impacts $n$ and $n\!\!+\!\!1$.
\item[] Assume $s\!>\!0$ and $0\!<\!\varepsilon\!<\!1$.
\end{enumerate}

As the proof of Theorem \ref{SuperProp} will be constructive, we select an initial condition $(x_{1}(0),v_{1}(0),x_{2}(0),v_{2}(0))=(x_{10},v_{10},x_{20},v_{20})$ such that
\begin{equation} \label{IC}
\left\{ \begin{array}{c}
x_{10}=s \\
v_{10}<0 \\
x_{20}>\tilde{x}_{2} \\
v_{20}<0
\end{array} \right.  
\end{equation}
Also, select a point $ \barra $ such that $\til<\barra<x_{20}$. 

Note that 
\begin{equation}
0<x_{2 eq}<\til<\barra<x_{20}
\end{equation}

Before turning to the actual proof of Theorem \ref{SuperProp},
 we must prove a series of lemmas.
We start by proving that lemmas \ref{f0} through \ref{f2bar} are true 
{\em while $x_{2}(t) \geq \barra$ holds}.

\begin{Lemma} \label{f0}
$0<F_{1}^{min}\leq F_{1}(t)$ when the system is under COD, where $F_{1}^{min}:=-(k+k_{12})s+k_{12} \barra$. 
\end{Lemma}

{\bf Proof:}
Under COD, the configuration of mass one and mass two that minimizes $F_{1}$ under the constraint 
$\left( x_{1}\leq s, x_{2} \geq \barra \right)$ is $\left( x_{1}=s, x_{2}=\barra \right)$.
In this configuration $F_{1}=-(k+k_{12})s+k_{12} \barra$ which is strictly larger than zero
since $\barra>\til$.$\Box$

\begin{Lemma} \label{f2<0}
$F_{2}(t)<0$ and $v_{2}(t)\leq v_{20}<0$.
\end{Lemma}

{\bf Proof:}
$F_{2}=0$ in the configuration $\left( x_{1},x_{2} \right)=\left( s, x_{2eq} \right)$. 
Since $x_{2}(t) \geq \bar{x}_{2}>x_{2eq}$ and 
$x_{1} \leq s$, the equations of motion guarantee $F_{2}(t)<0$.
Then $F_{2}(t)<0$ and $v_{20}<0$ imply $v_{2}(t) \leq v_{20}<0$.$\Box$

\begin{Lemma} \label{nocds}
$x_{1}(t)=s \Rightarrow v_{1}(t) \neq 0$.
\end{Lemma}

{\bf Proof:}
For mass one to reach the stop with zero velocity it would have to be decelerating as it 
approached the stop. This never happens since by Lemma \ref{f0}
 the mass always accelerates towards the stop.$\Box$
\\

Lemma \ref{nocds} is important since it rules out the possibility of CD phases while 
$x_{2}(t) \geq \barra$ holds. In other words, every time mass one reaches the stop there will
be a rebound. 
This fact will be assumed when deriving some of the lemmas that follow.

In the lemmas that follow we initialize $v_{out_{0}}$ with $v_{10}$.

\begin{Lemma}  \label{vinvout}
$\left| v_{in_{n+1}} \right| \leq \left| v_{out_{n}} \right|$.
\end{Lemma}

{\bf Proof:}
Lemma \ref{f0} implies that mass one decelerates as it moves 
monotonically from $s$ to some turning point $a$ and 
then accelerates as it moves monotonically from $a$ back to $s$, where the next rebound
occurs.  

 Now, $v_{2}(t)\leq v_{20}<0$ (Lemma \ref{f2<0}) implies that, for each $x \in 
\left( a,s \right)$, $F_{1}$ at $x$ is smaller when  mass one is
going back to the stop than when it was coming out of the stop.
 Hence, the magnitude of the work done 
between $s$ and $a$ to bring the mass from $\left| v_{out_{n}} \right|$
to rest is larger than the magnitude of the work done 
between $a$ and $s$ accelerating the mass from $0$ to $\left| v_{in_{n+1}} \right|$. 
Therefore the kinetic energy of mass one just before the $n\!\!+\!\!1$th rebound is less 
than its kinetic energy just after the $n$th rebound.$\Box$

\begin{Lemma} \label{voutvout}
$\left| v_{out_{n+1}} \right| \leq \varepsilon \left| v_{out_{n}} \right|$.
\end{Lemma}

{\bf Proof:}
Combine lemma \ref{vinvout} and the rebound law 
$v_{out_{n}}=-\varepsilon v_{in_{n}}$.$\Box$

\begin{Lemma}  \label{x1bar}
$x_{1}(t) \geq \bar{x}_{1} $ where $ \bar{x}_{1}:=s-\frac{v_{10}^{2}}{2 F_{1}^{min}}$.
\end{Lemma}

{\bf Proof:}
By lemma \ref{voutvout} the maximum speed of mass one out of a rebound is
$\left| v_{out_{0}}\right|:=\left| v_{10}\right|$.
 Now, assume mass one is acted throughout by the minimum possible force, $F_{1}^{min}$,
 as given in Lemma \ref{f0}.
 This assumption yields the stated $\bar{x}_{1}$ lower bound on $x_{1}(t)$.$\Box$

\begin{Lemma} \label{f2bar}
${F}_{2}^{min} \leq F_{2}(t)$ where ${F}_{2}^{min} :=-\left( k+k_{12} \right) x_{20}+k_{12} 
\bar{x}_{1}$.
\end{Lemma}

{\bf Proof:}
 By lemmas \ref{f2<0} and \ref{x1bar}, $x_{1}(t)\geq \bar{x}_{1}$ and
$x_{2}(t)\leq x_{20}$.
 The most negative value of $F_{2}(t)$ occurs when $m_2$ is at its rightmost position
and mass one is at its leftmost position.
 Therefore, ${F}_{2}^{min}$, the value of $F_{2}$ in the configuration 
$\left( x_{1},x_{2}\right) = \left( \bar{x}_{1},x_{20}\right)$, is a valid lowerbound on $F_{2}(t)$.
$\Box$
\\

This ends the set of lemmas that hold while $x_{2}(t) \geq \barra$.

\begin{Lemma} \label{preliminary}.

If 

\begin{enumerate}
\item The $n$th rebound occurred.
\item The inequality $x_{2}(t) \geq \barra$ holds throughout up to a time $\frac{2 
\left| v_{out_{n}} \right| }{F_{1}^{min}}$ after the $n$th rebound
\end{enumerate}
Then

\begin{enumerate}
\item[] The $n\!\!+\!\!1$st rebound is guaranteed to occur within that time interval, i.e.,
 $\Delta t_{n} \leq  \frac{2 \left| v_{out_{n}} \right| }{F_{1}^{min}}$.
\end{enumerate}
\end{Lemma}

{\bf Proof:}
This proof is by contradiction.
 Assume mass one has not yet reached the stop at the end of the time interval of length
 $\frac{2 \left| v_{out_{n}} \right| }{F_{1}^{min}}$. Then, by Lemma \ref{f0},
 $0<F_{1}^{min}\leq F_{1}(t)$ throughout the time interval of length 
$\frac{2 \left| v_{out_{n}} \right| }{F_{1}^{min}}$ following the $n$th rebound.
Therefore, at the end of this time interval, mass one must be moving again towards the stop
 with speed $\left| v_{1} \right| \geq \left| v_{out_{n}} \right|$.

Now, let $x_{1}^{*}<s$ be the position of mass one at the end of this time interval.
 Noting that $x_{1}^{*}$ is strictly less than $s$, 
 an argument parallel to the one used in Lemma \ref{vinvout} yields that
$\left| v_{1}\right|$  at $x_{1}^{*}$ must be strictly less than
$\left| v_{out_{n}}\right|$.
Hence we get a contradiction.$\Box$

\begin{Lemma} \label{grandiosa}
.

If
\begin{enumerate}

\item[] $x_{2}(t) \geq \barra$ holds for $t \in \left[ 0,
\sum_{i=0}^{n}\left( \frac{2 \varepsilon^{i}\left| v_{10} \right|}{F_{1}^{min}} \right) \right]$
\end{enumerate}
Then
\begin{enumerate}
\item[] the $n\!\!+\!\!1$th rebound occurs at a time $t\leq
\sum_{i=0}^{n}\left( \frac{2 \varepsilon^{i} \left| v_{10} \right|}{F_{1}^{min}} \right)$.
\end{enumerate}
\end{Lemma}

{\bf Proof:}
Follows from Lemma \ref{preliminary} and Lemma \ref{voutvout} using induction.
$\Box$
\\

We are now ready to prove Theorem \ref{SuperProp}.
\\

{\bf Proof of Theorem \ref{SuperProp}:}
Let T be defined as the time it takes mass two to reach \barra .
We would like to bound T below by $T_{1}$:
\begin{equation} 
0<T_{1}\leq T \label{basicT1}
\end {equation}
By Lemma \ref{f2bar}, this can be done by assuming $F_{2}(t)=F_{2}^{min}$
while $x_{2}(t)\geq \barra$.
Then $T{_1}$ is just the (positive) solution to the kinematic equation
\begin{equation} \label{T1equation}
\barra-x_{20}=v_{20}T_{1}+\frac{1}{2} F_{2}^{min}T_{1}^{2}
\end{equation}
Note that $T_{1}>0$ as asserted because both $F_{2}^{min}$ and $v_{20}$ are finite
 and $\barra<x_{20}$.
\\

Now, suppose, instead of $v_{1}(0)=v_{10}$, we had picked an initial condition 
$v_{1}(0)=v_{10}^{new}$, where $v_{10}<v_{10}^{new}<0$.

Looking at Lemma \ref{x1bar}, one sees that our old $\bar{x}_{1}$ lower bound on
$x_{1}(t)$ is still valid.
Likewise for the $F_{2}^{min}$ lower bound on $F_{2}(t)$ in Lemma \ref{f2bar}.
Therefore, $T_{1}$ calculated in (\ref{T1equation}) is also still a valid lower bound on the 
time $T^{new}$ it now takes mass two to reach \barra, i.e.,
\begin{equation}
0<T_{1}\leq T^{new} \label{T1new}
\end{equation}
\\
Noting that $\sum_{i=0}^{\infty}\left( \frac{2 \varepsilon^{i}\left| v_{10}^{new} \right|}{F_{1}^{min}} 
\right)
=\frac{2 \left| v_{10}^{new} \right|}{F_{1}^{min}} \frac{1}{1-\varepsilon}$
, it is clear that we can pick a $v_{10}^{new}$ value 
such that $\sum_{i=0}^{\infty}\left( \frac{2 \varepsilon^{i}\left| v_{10}^{new} \right|}{F_{1}^{min}} \right)
<T_{1}$ is also satisfied.
Then, by Lemma \ref{grandiosa}, it follows that mass one undergoes an infinite 
number of rebounds in the finite time 
$\frac{2 \left| v_{10}^{new} \right|}{F_{1}^{min}} \frac{1}{1-\varepsilon}$.
\\

We have shown that for any
 $x_{20}>\tilde{x}_{2}$, $v_{20}<0$, we can find a $v_{10}^{new}<0$ such that the initial condition  
$(x_{1}(0),v_{1}(0),x_{2}(0),v_{2}(0))=(s,v_{10}^{new},x_{20},v_{20})$
leads to infinite rebounds in finite time.
 Now, if in any such $(s,v_{10}^{new},x_{20},v_{20})$ initial condition we replace $v_{10}^{new}$ by
$v_{10}^{*}$, where $v_{10}^{new}<v_{10}^{*}<0$, we still have an initial condition that leads to
infinite rebounds in finite time.
This is so because:
\begin{itemize}
\item $T_{1}$ is still a valid lower bound for the time it now takes mass two to reach \barra.  
\item $
 \sum_{i=0}^{\infty}\left( \frac{2 \varepsilon^{i} \left| v_{10}^{*} \right|}{F_{0}} \right)<
 \sum_{i=0}^{\infty}\left( \frac{2 \varepsilon^{i} \left| v_{10}^{new} \right|}{F_{0}} \right)$.
\end{itemize}

Therefore we have a 3-D region in the 4-D phase space that leads to infinite rebounds in 
finite time.
Noting that 
\begin{itemize}
\item This region is contained in the hypersurface $\{x_{1}\!=\!s\}$.
\item $v_{1}<0$ for all the points  in this region.
\end{itemize}
we conclude that orbits through this region evolve transversely to it.
Hence the region of phase space that leads to infinite rebounds in a finite time has nonzero measure in the phase space.$\Box$

\subsection{Continuation of orbits that lead to infinite switching} \label{analyticalcont}

In order for an orbit that leads to infinite switching to be well defined for all time,
 we postulate that orbit to continue from the point approached by the infinite rebounds.
We call this a {\em reset} of the orbit.

\begin{Def}[Orbit Reset] \label{orbitresets}
Let ${\bf x}( t;{\bf x_{0}})$ be an orbit that leads to infinite switching between
COD dynamics and R dynamics as
$t\longrightarrow t^{*}$.
Assume that ${\bf x}( t;{\bf x_{0}}) \longrightarrow {\bf x^{*}}$ as $t\longrightarrow t^{*}$.

For $t\geq t^{*}$ we define
 ${\bf x}( t;{\bf x_{0}})={\bf x}( t-t^{*};{\bf x^{*}})$.
We call ${\bf x}^{*}$ a {\em reset point}.
\end{Def}

For this to be a valid procedure, one must show that a limit point is indeed 
always approached under this form of infinite switching.

\begin{Prop} \label{LIMIT}
Let ${\bf x}( t,{\bf x_{0}})$ be an orbit that leads to infinite switching 
between COD dynamics and R dynamics as
$t\longrightarrow t^{*}$.
 Then, ${\bf x}( t,{\bf x_{0}}) \longrightarrow {\bf x^{*}}$ as 
$t\longrightarrow t^{*}$, where ${\bf x^{*}}=(s,0,x_{2}^{*},v_{2}^{*})$.
\end{Prop}

Observe that, from the boundedness of energy, it follows that 
$x_1(t), v_{1}(t), x_{2}(t)$ and $v_{2}(t)$ are bounded for $t\in [0,t^{*})$. From this, in turn,
 it follows that $F_{2}(t)$ and $F_{1}(t)$, except at rebounds, are also bounded for $t\in [0,t^{*})$.

Now we prove Proposition \label{limit} by proving a series of lemmas.

\begin{Lemma} \label{lemmax1}
Let ${\bf x}( t,{\bf x_{0}})$ be an orbit that leads to infinite rebounds as
$t\longrightarrow t^{*}$.
Then $x_{1}(t)\longrightarrow s$ as $t\longrightarrow t^{*}$.
\end{Lemma}
{\bf Proof:}
Since there are infinite rebounds as $t\rightarrow t^{*}$, one can find a time
arbitrarily close to $t^{*}$ at which $x_{1}(t)=s$. Together with the
boundedness of $v_{1}(t)$, this implies that $x_{1}(t)\longrightarrow s$ as $t\longrightarrow t^{*}$.

\begin{Lemma} \label{lemmav1}
Let ${\bf x}( t,{\bf x_{0}})$ be an orbit that leads to infinite rebounds as
$t\longrightarrow t^{*}$.
Then $v_{1}(t)\longrightarrow 0$ as $t\longrightarrow t^{*}$.
\end{Lemma}
{\bf Proof:}
First we prove that
 $Lim_{t\rightarrow t^{*}}$ Inf $v_{1}(t) \geq 0$.
 Assume not. That is, assume that $\exists\epsilon>0$ s.t.
 $\forall t_{1}, \exists t_{2}$ with $t_{1}<t_{2}<t^{*}$ and $v_{1}(t_{2}) \leq -\epsilon$.
 Now, because $F_{1}(t)$ is bounded under COD, it is possible to bound away 
from 0 the time it takes for mass one to go from a velocity equal or more negative
than $-\epsilon$ to rest. This bounds acts also as lower  
a bound on the time it takes for mass one to reach the stop again.
 But we can find a $t_{2}$ arbitrarily close to $t^{*}$ satisfying $v_{1}(t_{2}) \leq -\epsilon$.
This contradicts the existence of infinite rebounds as $t\longrightarrow t^{*}$.

Now we prove that $Lim_{t\rightarrow t^{*}} $Sup $v_{1}(t) \leq 0$.
 Assuming otherwise, together with 
$x_{1}(t)\rightarrow s$ as $t\rightarrow t^{*}$ and the boundedness of
$F_{1}(t)$ under COD dynamics, implies that there exists an $\epsilon>0$
such that, arbitrarily close to $t^{*}$, one can find a rebound with 
a velocity larger than or equal to $\epsilon$.
 Considering the velocities after those rebounds, one see that this violates 
$Lim_{t\rightarrow t^{*}}$ Inf $v_{1}(t) \geq 0$.$\Box$

\begin{Lemma} \label{lemmax2v2}
Let ${\bf x}( t,{\bf x_{0}})$ be an orbit that leads to infinite rebounds as
$t\longrightarrow t^{*}$.
Then there exist values $x_{2}^{*}$ and $v_{2}^{*}$ such that 
$x_{2}(t)\longrightarrow x_{2}^{*}$ as $t\longrightarrow t^{*}$
and $v_{2}(t)\longrightarrow v_{2}^{*}$ as $t\longrightarrow t^{*}$.
\end{Lemma}
{\bf Proof:}
$v_{2}(t)$ has a limit since $F_{2}(t)$, i.e., the derivative of $v_{2}(t)$,
is bounded for $t\in [0,t^{*})$.
$x_{2}(t)$ has a limit since $v_{2}(t)$, i.e., the derivative of $x_{2}(t)$,
 is bounded for $t\in [0,t^{*})$.$\Box$
\\

The proof of Proposition \ref{LIMIT} follows at once from lemmas \ref{lemmax1},
 \ref{lemmav1} and \ref{lemmax2v2}.
\\

We have therefore solved the problem of infinite switching between
{\em COD} dynamics
and {\em R} dynamics in a finite time. 
Next, we would like to rule out any form of infinite switching in finite time involving 
CD dynamics segments. This will follow as a corollary to the proposition below.

\begin{Prop} \label{CDtimebounded}
It is possible to bound away from zero the time elapsed between any two {\em CD} dynamics segments.
\end{Prop}
{\bf Proof:}
First note that $F_{2}(t)<0$ for $x_{2}(t) \in (x_{2eq},\tilde{x}_{2}]$ and that this
is valid under all governing dynamics and regardless of the position of mass one.

Now, from chart~\ref{bigchart}, we know that
 at the end of a CD segment $x_{2}(t)=\tilde{x}_{2}$ and $v_{2}(t)<0$.
Therefore, after a CD segment,
 mass two will move monotonically from $\tilde{x}_{2}$ to $x_{2eq}$.
Since $v_{2}(t)$ is bounded, we can bound away from zero, say by $T$, the time it takes
mass two to go from $\tilde{x}_{2}$ to $x_{2eq}$.
 Note that this analysis still holds even if orbit resets due to infinite switching between
COD and R dynamics occurs in the process. This is the case because an orbit reset maintains
the continuity of $x_{2}(t)$ and $v_{2}(t)$.  

Now, the next CD segment can be joined either following a COD segment or following an
 orbit reset due to infinite switching between {\em COD} dynamics and {\em R} dynamics.
In the first case the CD segment can be joined only at a point where 
mass two is at $\tilde{x}_{2}$ again.
In the case of an orbit reset at $t^{*}$
it can only be joined if $lim_{t \rightarrow t^{*}}x_{2}(t) \geq \tilde{x}_{2}$.
In either case $T$ is clearly a valid lower bound on the time elapsed between two
CD segments.
$\Box$

\begin{Cor}
Infinite switching in a finite time involving CD dynamics is not possible.
\end{Cor}

In other words, infinite switching in finite time occurs only between
COD dynamics and R dynamics. 
Definition \ref{orbitresets} ({\em Orbit Reset}) is therefore all that is needed as far as 
defining continuation of orbits that undergo infinite switching between 
different dynamics in a finite time. 

The last step towards establishing forward existence and uniqueness is to
 show that there can be no infinite orbit resets themselves 
in a finite time. We start with a lemma about what are possible {\em reset points}.

\begin{Lemma} \label{reset point}
Let ${\bf x}( t;{\bf x_{0}})$ be an orbit that leads to infinite rebounds as
$t\longrightarrow t^{*}$.
Let ${\bf x}(t) \rightarrow x^{*}$ as $t\longrightarrow t^{*}$. 
Then $x^{*}$ must be of the form $(s,0,x_{2}^{*}, v_{2}^{*})$, where $x_{2}^{*}$ is equal or
greater than $\tilde{x}_{2}$.
\end{Lemma} 
{\bf Proof:}
That $x_{1}^{*}=s$ and $v_{1}^{*}=0$ is shown in Proposition \ref{LIMIT}.
Now, assume $x_{2}^{*}<\tilde{x}_{2}$.
Then, $x_{2}(t) \rightarrow x_{2}^{*}$ as $t\longrightarrow t^{*}$
implies that it is possible to find a constant $c_{1}$ and a time $t_{1}$ before $t^{*}$ 
such that, 
for $t\in (t_{1},t^{*})$,  $x_{2}(t)<c_{1}<\tilde x_{2}$.
In turn this means one can find a nonempty interval $(c_{2},s)$ such that, 
for all 
$t\in (t_{1},t^{*})$, when $x_{1}(t)$ is in $(c_{2},s)$, $F_{1}$ is
negative.

Now consider any rebound occuring at a time $t\in (t_{1},t^{*})$.
By the above analysis, 
after the rebound, mass one will have to leave the region $(c_{2},s)$ before it can   
turn around and return for another rebound.
 Since $v_{1}(t)$ is bounded, one can bound away from zero the time it takes 
for mass one to go from $s$ to $c_{2}$.
This contradicts the existence of infinite rebounds as $t\rightarrow t^{*}$.  
$\Box$

\begin{Cor} \label{afterthereset}
After an orbit reset one of the following two types of segments is joined:
\begin{enumerate}
\item A CD segment (this is the case if $x_{2}^{*}>\tilde{x}_{2}$ or 
$(x_{2}^{*}=\tilde{x}_{2}$ and $v_{2}^{*}>0)$).
\item A COD segment starting at a point of the form ${\bf x^{*}}=(s,0,\tilde{x}_{2},v_{2}^{*})$
where $v_{2}^{*}\leq 0$.
\end{enumerate} 
\end{Cor}

\begin{Prop}
It is not possible to have infinite {\em orbit resets} in a finite time.
\end{Prop}
{\bf Proof:}
Each orbit reset is followed by a segment of one of the two types outlined in the
corollary above.
Let us look at each possibility and its implications separately.
Consider an orbit reset that leads to a COD segment starting at a point of the form 
${\bf x^{*}}=(s,0,\tilde{x}_{2},v_{2}^{*})$, where $v_{2}^{*}\leq 0$.
First note that $F_{2}(t)<0$ for $x_{2}(t) \in (x_{2eq},\tilde{x}_{2}]$ and that this
is valid under all governing dynamics and regardless of the position of mass one.
 Together with $x_{2}^{*}=\tilde{x}_{2}$ and
$v_{2}^{*}\leq 0$, this allows us to conclude that mass two will be moving monotonically from
 $\tilde{x}_{2}$ to $x_{2eq}$.
 Since $v_{2}(t)$ is bounded, we can bound away from zero, say by $T$, the time it takes
mass two to go from $\tilde{x}_{2}$ to $x_{2eq}$.
 As explained in the proof of proposition \ref{CDtimebounded}, this 
holds even if orbit resets occur in the meantime. 
 Now, the next orbit reset leading to a segment of this type 
can only occur when the limit of $x_{2}(t)$ is again $\tilde{x}_{2}$.
 Therefore T is also a bound on the time elapsed before the next orbit reset which leads to 
a segment of this type. 
Thus, in a finite time, there can only be finitely many orbit resets that lead to segments of this type.

Now, consider an orbit reset that leads to a CD segment. By proposition~\ref{CDtimebounded}, the time
elapsed between two CD segments is bounded away from zero. Therefore, so is the time elapsed between
 two orbit resets that lead to CD segments. Hence, in a finite time, there can only be finitely
 many orbit resets that lead to a CD segment.

Putting the two cases together, one concludes that in a finite time there can only
be finitely many orbit resets.$\Box$
\\

We conclude that the addition of postulate \ref{orbitresets} to the mathematical model ensures 
forward existence and uniqueness of solutions for all time.

Note that, however,
 backwards uniqueness is broken by orbit resets, since when an orbit undergoes a reset
 it joins a regular orbit of the system. In fact, orbit resets are the only way in which backwards uniqueness
 is broken in the system.

\section{The Dynamics of the System}

In this second part of the paper we analyze the dynamics of the system. Rather than providing a long and exhaustive characterization, we highlight some of the more interesting features of the system. However, for completeness, we first state a general theorem concerning the forward limit set of the system.

\begin{Theo}
 The forward limit set of the system is the set of orbits that are
governed exclusively by coupled oscillator dynamics and contact dynamics,
in other words, the set of orbits that do not go through any rebound.
\end{Theo}

Since the theorem is quite plausible and the proof is long, we ommit the proof. It relies on a LaSalle standard type of argument\footnote{See Khalil \cite{Khalil}.}. We just point out that, to use such an argument, invariance of the forward limit set must first be proved. Note that this is not immediate, because for a discontinuous system the forward limit set is not necessarily invariant, as the example below illustrates.

\begin{Example}
A discontinuous dynamical system with a non-invariant forward limit set.
\end{Example}
\hspace{3cm} $\dot{x}=f(x)$, where
$ f(x)= \left\{ \begin{array}{ll}
-x & \mbox{if $x<0$} \\
1 & \mbox{if $x \geq 0$} \\
\end{array}
\right. $
\\

Clearly, the only point in the forward limit set of this system is the point $x=0$.
However, the forward limit set is not invariant since $x=0$ is not a fixed point of the dynamics. 
$\Box$
\\

\subsection{Periodic hybrid orbits} \label{PHO section}

In this section we study orbits which contain both CD segments and COD segments but no rebounds, that is, orbits where mass one alternates finite periods against the stop with periods in which it oscillates together with mass two under coupled oscillator dynamics. Since mass one must avoid any rebounds, such orbits might seem unlikely. However they do exist. It turns out that all such orbits are periodic and consist of exactly two and only two segments, one governed by the contact dynamics, the other by the coupled oscillator dynamics. We call them 2-phase periodic hybrid orbits. We start by showing why these 2-phase periodic hybrid orbits are the only possible orbits made of just CD and COD segments.
\\

Let us track such an hypothetical hybrid orbit, starting with a CD segment. First note that, following a CD segment, a COD segment can only be joined at a point in the set $\{s,0,\tilde{x}_{2},v_{2}\!<0\}$. This is so, since these
are the exiting points of the CD region (see chart~\ref{bigchart}).
Let the COD segment be joined at the point $(s,0,\tilde{x}_{2},{v}^{*}_{2})$
, where $v_{2}^{*}$ is a negative number. Now, this point, when evolved under COD in
 backwards time, comes from the $\{x_{1}\!>\!s\}$ region. This means that, whether the COD orbit is periodic or quasiperiodic, if the system stays long enough under coupled oscillator dynamics, the orbit will return to the $\{x_{1}\!>\!s\}$ region. But we know this can never happen, so eventually a rebound or a CD segment must follow that COD segment. Assume a CD segment rather than a rebound follows the COD segment.
 This next CD segment can be joined only at a point on the entrance surface of the CD region/exit surface of the COD region. From chart~\ref{bigchart}, this must be a point of the form $(s,0,\tilde{x}_{2},{v}^{**}_{2})$, where $v_{2}^{**}$ is a positive number.
 Furthermore, because energy was conserved during the preceeding COD segment,
 we must have that ${v}^{**}_{2}=-{v}^{*}_{2}$.
 Next, during the CD segment mass two oscillates as a harmonic oscillator for
 a fraction of a complete cycle. If the CD segment started at 
$(s,0,\tilde{x}_{2},-{v}^{*}_{2})$, it must end at $(s,0,\tilde{x}_{2},{v}^{*}_{2})$
 (see Figure \ref{hypersurface}).
 At this instant the orbit as exactly the same initial conditions as at the beginning of the description. Therefore everything simply repeats. We conclude that
\\

{\em The periodic orbits described above are the only possible orbits that contain both
 Coupled Oscillator Dynamics and Contact Dynamics and no Rebounds at all. Conservation of energy
 precludes the possibility of more complicated orbits, periodic or otherwise.}
\\

We now turn to the task of explicitly constructing 2-phase periodic hybrid solutions analytically.
\\

Let the solution to the COD segment of the hybrid orbit 
start at time $t=0$ and be expressed as

\begin{equation} \label{CODsol}
\left\{ \begin{array}{l}
x_{1}(t)= \Re(\underline{J}_{1} e^{i\omega_{1}t}+\underline{J}_{2} e^{i\omega_{2}t})\\
x_{2}(t)= \Re(\underline{J}_{1}e^{i\omega_{1}t}-\underline{J}_{2} e^{i\omega_{2}t})\\
\end{array} \right.  
\end{equation}
where $\omega_{1}=\sqrt{k}$, $\omega_{2}=\sqrt{k+2k_{12}}$
and $\underline{J}_{1}$ and $\underline{J}_{2}$ are complex constants related to initial conditions by

\begin{equation} \label{ICforCOD}
\left\{ \begin{array}{l}
x_{1}(0)=\Re(\underline{J}_{1})+\Re(\underline{J}_{2})\\
v_{1}(0)=-\omega_{1}\Im(\underline{J}_{1})-\omega_{2}\Im(\underline{J}_{2})\\
x_{2}(0)=\Re(\underline{J}_{1})-\Re(\underline{J}_{2})\\
v_{2}(0)=-\omega_{1}\Im(\underline{J}_{1})+\omega_{2}\Im(\underline{J}_{2})\\
\end{array} \right.
\end{equation}
\\

 This is a segment of an hybrid orbit if and only if the following set of
five equations and one inequality are satisfied for some integer $n\in N$.
\begin{equation}\label{16}
s=\Re(\underline{J}_{1})+\Re(\underline{J}_{2})
\end{equation}
\begin{equation}\label{17}
0=-\omega_{1}\Im(\underline{J}_{1})-\omega_{2}\Im(\underline{J}_{2})
\end{equation}
\begin{equation}\label{18}
\tilde{x}_{2}=\Re(\underline{J}_{1})-\Re(\underline{J}_{2})
\end{equation}
\begin{equation}\label{19}
{v}_{2}(0)=-\omega_{1}\Im(\underline{J}_{1})+\omega_{2}\Im(\underline{J}_{2})
\end{equation}
\begin{equation}\label{20}
\frac{2\pi-2 \arctan\frac{\Im(\underline{J}_{1})}{\Re(\underline{J}_{1})}}
{2\pi-2 \arctan\frac{\Im(\underline{J}_{2})}{\Re(\underline{J}_{2})}+2\pi n}
=\frac{\omega_{1}}{\omega_{2}}
\end{equation}
\begin{equation}\label{21}
x_{1}(t)\leq s  \mbox{ is satisfied }\forall t \in [0,t_{cod}], \mbox{ where }  
t_{cod}=\frac{2\pi-2 \arctan\frac{\Im(\underline{J}_{1})}{\Re(\underline{J}_{1})}}
{\omega_{1}},
\end{equation}
$t_{cod}$ being the duration of the COD phase of the hybrid orbit.
\\

(Note: the $\arctan$ range in these and all the subsequent formulas should be taken as $[0,\pi]$.)
\\

In the above set of equations, (\ref{16}) ensures that mass one is initially at the stop, (\ref{17}) ensures
 that its initial velocity is zero and (\ref{18}) guarantees that mass two is initially at $\tilde{x}_{2}$.
 Equation (\ref{19}) simply relates $v_{2}(0)$, the initial velocity of mass two, to the initial condition 
variables $\underline{J}_{1}$ and $\underline{J}_{2}$. No particular requirement is imposed on the value of
 $v_{2}(0)$. Equation (\ref{20}) guarantees that $x_{1}(t_{cod})\!=\!x_{1}(0)\!=\!s$, 
$v_{1}(t_{cod})\!=\!v_{1}(0)\!=\!0$, $x_{2}(t_{cod})\!=\!x_{2}(0)\!=\!\tilde{x}_{2}$ and  $v_{2}(t_{cod})\!=\!-v_{2}(0)$.
 Finally, inequality (\ref{21}) 
ensures that mass one stays to the left of the stop throughout the COD segment.
 The first five equations can be reduced to one by eliminating the complex variables 
$\underline{J}_{1}$ and $\underline{J}_{2}$. The resulting equation is transcendental involving
 the constants $k$, $k_{12}$ and $s$ and the required initial condition of mass two, $v_{2}(0)$.

\begin{equation}\label{transcendental}
\left\{ \begin{array}{l}
\frac{
2\pi-2\arctan \frac{-v_{2}(0)}{\sqrt k (2+\frac{k}{k_{12}})s}
}
{
2\pi +2n\pi-2\arctan \frac{-k_{12} v_{2}(0)}{s k\sqrt{k+2 k_{12}}}
}
=

\frac{\sqrt{k}}{\sqrt{k+2 k_{12}}}\\
x_{1}(t)\leq s :\ \forall t\in[0,t_{cod}],\mbox{ where }t_{cod}=
\frac{
2\pi-2\arctan\frac{-v_{2}(0)}{\sqrt{k}(2+\frac{k}{k_{12}})s}
}{\omega_{1}}\\

\end{array} \right.
\end{equation}
\\

So, for any integer $n$ for which one can find a value of $v_{2}(0)$ that satisfies 
(\ref{transcendental}), one has a valid COD segment of a 2-phase periodic hybrid solution.
 The constants $\underline{J}_{1}$ and $\underline{J}_{2}$ in the analytical solution (\ref{CODsol})
 of the COD segment are then
\begin{equation}
\left\{ \begin{array}{l}
\underline{J}_{1}=\frac{1}{2}(2+\frac{k}{k_{12}})s+i\frac{-v_{2}(0)}{2\sqrt{k}}\\
\underline{J}_{2}=\frac{-k s}{2 k_{12}}+i\frac{v_{2}(0)}{2\sqrt{k+2 k_{12}}}\\
\end{array} \right.
\end{equation}

and this solution is valid for $t\in [0,t_{cod}]$, the duration of the COD segment.
\\

The analytical solution to the CD segment of the orbit is
\begin{equation}
\left\{ \begin{array}{l}
x_{1}(t)=s\\
x_{2}(t)=x_{2eq}+\Re(\underline{C}e^{i\omega_{shm}(t-t_{cod})})\\
\end{array} \right.
\end{equation}

where $\omega_{shm}=\!\sqrt{k+k_{12}}$,   $\:\underline{C}=(\frac{k+k_{12}}{k_{12}})s-x_{2eq}+i\frac{v_{2}(0)}
{\omega_{shm}}$
\\

and this solution is valid for 
$$t\in[t_{cod},t_{cod}+t_{cd}],\mbox{ where }t_{cd}=
\frac{
2\pi-2\arctan
\frac{\frac{v_{2}(0)}{\omega_{shm}}}{\frac{k+k_{12}}{k_{12}}s}
}
{\omega_{shm}}$$
\\
The constant $\underline{C}$ was chosen so that the boundary conditions 
\begin{equation}
\left\{ \begin{array}{l}
x_{2}(t_{cod})=\tilde{x}_{2}\\
v_{2}(t_{cod})=-v_{2}(0)\\
\end{array} \right.
\end{equation}
between the COD and CD segments are met.
\\

\begin{figure}[!h] 
\epsfysize=2in
\centerline{\epsfbox{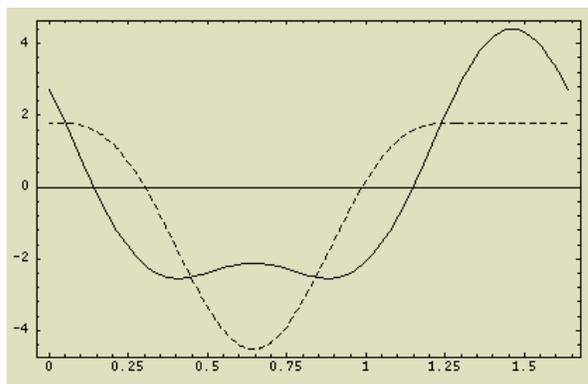}}
\caption{A 2-phase periodic hybrid orbit.}  \label{hybrid figure}
\end{figure}

An example of a 2-phase periodic hybrid orbit in a system with
 spring constants $k\!=\!13,\: k_{12}\!=\!25$ and the stop placed at $s\!=\!1.8$ is shown in Figure
 \ref{hybrid figure}.
 The dotted line represents the trajectory of mass one.
 The continuous line represents the trajectory of mass two.
 In this case $v_{2}(0)\!=\!-17.47$ satisfied the transcendental equation in (\ref{transcendental}) for $n\!=\!1$.
 The times $t_{cod}\!=\!1.289$ and $t_{cd}\!=\!0.26$, so the period of this hybrid orbit is
$t_{cod}+t_{cd}\!=\!1.549$.

\vspace{.1in}

\subsection{Dynamics of action angle coordinates and simulations} \label{Action section}

In this section we analyze the dynamics in the action-angle coordinates associated with the COD.
 This allows a simple geometrical way of interpreting and visualizing the system dynamics.
 The analysis in this section concerns only the case in which the COD orbits are quasiperiodic.
  Assume first that there is no stop. 
  Parametrize the tori in which COD orbits live by ($I_{1}$, $\!I_{2}$), the value the COD action variables take on those tori.
   A COD orbit expressed as in (\ref{CODsol}) lives in the torus where the action variables take the values $I_{1}=|\underline{J}_{1}|$ and $I_{2}=|\underline{J}_{2}|$.
   
\begin{Def}
.
\\
$({I}_{1},{I}_{2})$ Torus $\equiv
\left\{ \begin{array}{l}
 (x_{1},v_{1},x_{2},v_{2})\in R^{4}:
\left( \begin{array}{l}
I_{1}=\sqrt {(\frac{x_{1}+x_{2}}{2})^2+(\frac{v_{1}+v_{2}}{2 \omega_{1}})^2}\\
I_{2}=\sqrt {(\frac{x_{1}-x_{2}}{2})^2+(\frac{v_{1}-v_{2}}{2 \omega_{2}})^2}\\
\end{array} \right)
\end{array} \right\} $
\end{Def}

Referencing orbits by the torus in which they live captures most of the information, since quasiperiodic orbits in the same torus are qualitatively similar. Therefore we will now study the evolution of orbits solely in terms of of the evolution of their COD action coordinates. We call the space defined by the $I_{1}$ and $I_{2}$ action coordinates the action space. Naturally, without the stop, orbits never leave the torus they start in, i.e., they remain at a fixed point in the action space. Now we describe how the introduction of the stop changes this. The effect of introducing the stop is to divide the tori into three distinct groups:
\begin{enumerate}
\item Tori satisfying $I_1+I_2<s$.

These are tori that are a finite distance away from the $\{x_{1}\!=\!s\}$ surface. We call them {\em interior tori}. The orbits living in the interior tori are the ones where mass one stays always a finite nonzero distance away from the stop.

\item Tori satisfying $I_1+I_2=s$.

These are tori that are tangent to the $\{x_{1}\!=\!s\}$ surface. We call them {\em grazing tori}.
 Grazing tori are the tori that touch the hypersurface $\{x_{1}\!=\!s\}$ exactly at one point, of the form $(s,0,x_{2},0)$, where $x_{2}$ is between $-s$ and $s$. As the name indicates, on orbits living in grazing tori mass one comes arbitrarily close to the stop, on some of those orbits actually touching it with zero speed.

\item Tori satisfying $I_1+I_2>s$.
 
 These are tori that are cut by the $\{x_{1}\!=\!s\}$ surface.
 
 \end{enumerate}
 
\begin{figure}[h]
\epsfysize=2in
\centerline{\epsfbox{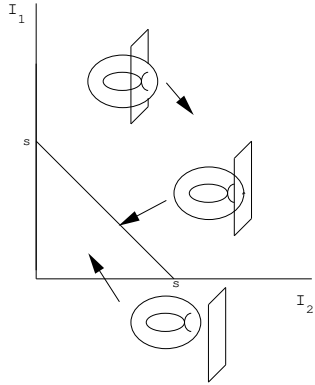}}
\caption{Action space.}  \label{Maria}
\end{figure}

  This division is shown in Figure \ref{Maria}. The tori on the diagonal line going from $(0,s)$ to $(s,0)$ are the grazing tori. The tori to the right of the line are cut by the stop, while the ones to the left of the line are the interior tori. 
\\
 
An orbit that starts in a torus that is cut by the stop winds around that torus until it intersects the surface where the torus is cut. At that point, physically, a rebound occurs. In the action space, the orbit simply jumps to another torus at a lower energy level. It then starts winding around the new torus until it again reaches the surface where the torus is cut, at which point the orbit jumps again to a torus of lower energy. The jumps are never to one of the interior tori since interior tori form an invariant set. Also, because of quasiperiodicity, an orbit in a torus which is cut by the stop cannot remain under COD indefinitely without intersecting the $\{x_{1}\!=\!s\}$ surface.
 Orbits in these tori are therefore guaranteed to eventually jump to a lower energy torus. 
The exception to this is the 2-phase periodic hybrid orbit. The COD segments of 2-phase periodic hybrid orbits {\em are} in tori that are cut by the stop. However, when a 2-phase periodic hybrid orbit reaches the $\{x_{1}\!=\!s\}$ surface, it joins a CD segment instead of undergoing a rebound.   
 Then, when the CD segment is over, the exact same COD segment that lead to it is rejoined.
 The 2-phase periodic hybrid orbits are therefore these very special orbits,
 the only that manage to survive in tori that are cut by the stop.
\\

\begin{figure}[h]
\epsfysize=2in
\centerline{\epsfbox{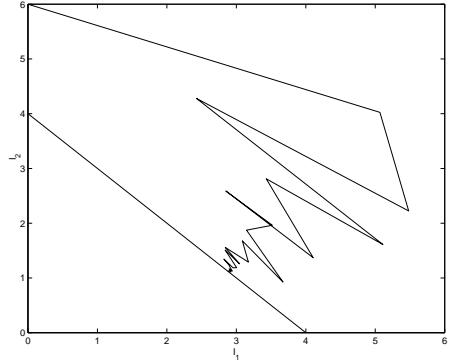}}
\caption{A trajectory in action space.}  \label{figurex00}
\end{figure}

\begin{figure}[h] 
\epsfysize=2in
\centerline{\epsfbox{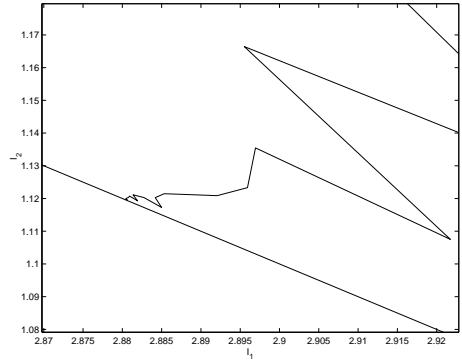}}
\caption{Zoom of Figure \ref{figurex00} near the grazing tori line.} \label{zoom}
\end{figure}

Figure \ref{figurex00} is a numerical simulation of the evolution of an orbit in the action
 space. The tori which the orbit visits have been connected by straight lines in order to 
allow one to discern the sequence in which the tori are visited.
 In this simulation the stop was placed at $s\!=\!4$, the coefficient of restitution was
 $\varepsilon\!=\!0.8$ and the spring constants were $k\!=\!5.3657$ and $k_{12}\!=\!1.5682$.
 The initial condition of the orbit shown is
 $(x_{1}(0),v_{1}(0),x_{2}(0),v_{2}(0))=(-6,0,6,0)$.
 As the orbit jumps from torus to torus, it asymptotically approaches the line of
 grazing tori.
 As it does so, the jumps become smaller and smaller. This can be seen in
Figure \ref{zoom}, a zoom of Figure \ref{figurex00} near the grazing tori line.
 Predicting which grazing torus, if any, the orbit will specifically approach, remains an open
 research question.
 
\begin{figure}[h] 
\epsfysize=5in
\centerline{\epsfbox{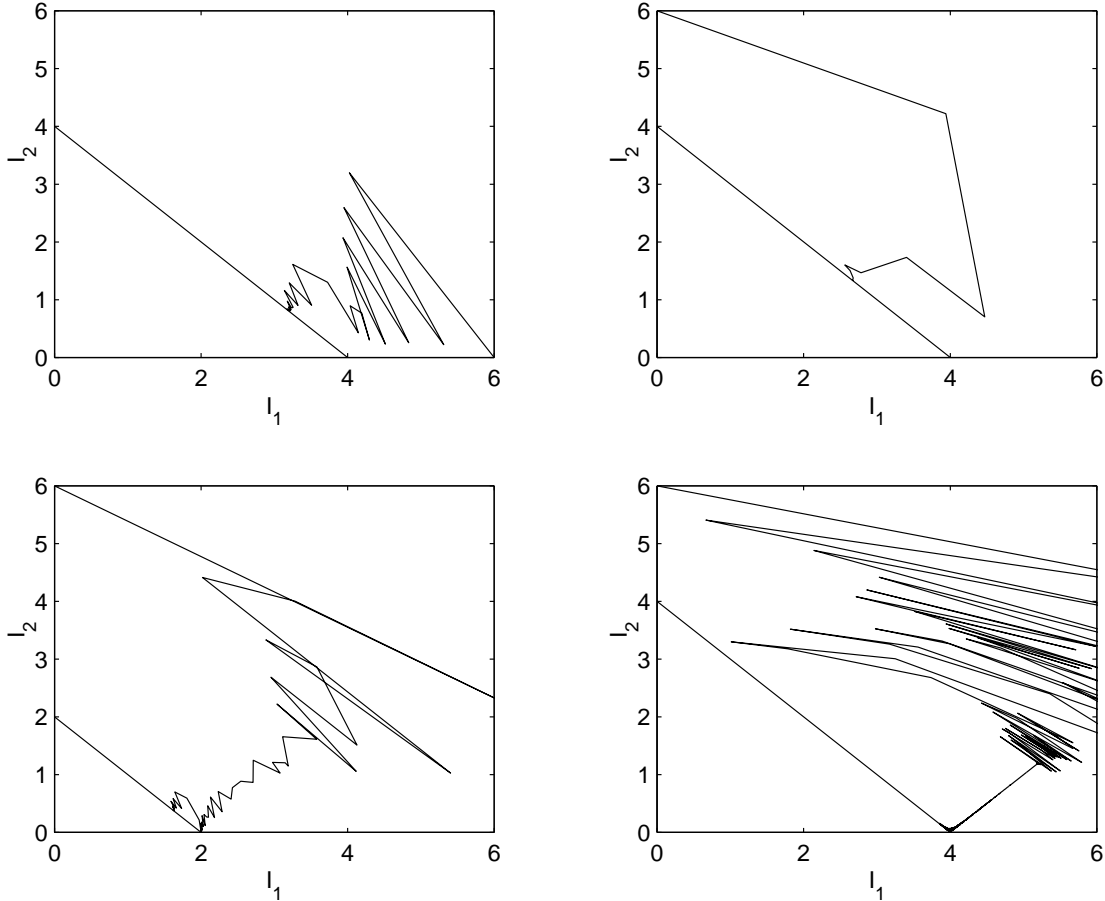}}
\caption{Trajectories under different parameters and initial conditions.} \label{showcasetori}
\end{figure}

Figure \ref{showcasetori} shows a few more simulations that illustrate
a range of the behaviors that can be observed. 
 On each simulation exactly one parameter was changed from the parameters used in the simulation in figure \ref{figurex00}.
In the top left picture, a different initial condition, $(x_{1}(0),v_{1}(0),x_{2}(0),v_{2}(0))=(6,0,6,0)$, was used.
 On the top right picture the coefficient of friction was reduced to $\varepsilon\!=\!0.4$.
 Hence, more energy is lost at each impact and within a small number of impacts the system comes very close to a grazing torus.
 On the bottom left picture, the stop was placed at $s\!=\!2$.
 Interestingly, this orbit seems at first to be approaching a torus near
 $(I_{1},I_{2})=(2,0)$
 but then suddenly, with two or three larger jumps, moves to 
the region near $(I_{1},I_{2})=(1.5,0.5)$.   
On the bottom right picture, the spring constants were changed to
 $k\!=\!2.5465$ and $k_{12}\!=\!3.9780$. The trajectory in this picture clearly undergoes three qualitatively distinct phases. First there are jumps between far apart tori. Then, abruptly, the trajetory changes to very small jumps which gradually lead the system to the region next to $({I}_{1},{I}_{2})=(4,0)$.
 Finally, once close to $({I}_{1},{I}_{2})=(4,0)$, the trajectory slowly creeps up, parallel to the grazing tori line.

We point out that this section's analysis is not pertinent when the COD orbits are periodic rather than quasiperiodic.
 In that case, there exist COD orbits that never intersect the $\{x_{1}\!=\!s\}$ surface even though they live in tori that are cut by the stop.

\section{Conclusions}
In this paper we studied a coupled oscillator system with inelastic impacts. We started by showing that applying the standard model in which the system switches between the different types of governing dynamics results in a fundamental problem, namely the existence of a nonzero measure set in the phase space where infinite rebounds occur in a finite time. This problem was solved by continuing such orbits from the limit point approached by the infinite rebounds. Then, looking at the system dynamics, we found a particularly interesting type of periodic behavior, the 2-phase periodic hybrid orbits, which we described analytically. Finally, we presented some numerical simulation results and a way of visualizing and interpreting them, based on action-angle coordinates.

This paper highlighted two aspects of hybrid systems. First, it stressed the care that must be taken to ensure a well defined mathematical model for the hybrid system. Secondly, it demonstrated that hybrid systems, even elementary ones, can possess quite interesting and unusual dynamics. We hope this will serve as a useful step towards further theoretical studies of multi-degree of freedom hybrid systems.

\section{Acknowledgements}
Andr\'{e} Valente was supported by a fellowship from the Portuguese Ministry of Science and Technology. Igor Mezi\'{c} was partially supported by a Sloan fellowship. The work of Harris McClamroch was partially supported by NSF Grant ECS-9906018.

\end{document}